\documentclass[runningheads]{llncs}
\usepackage{graphicx}
\usepackage{amsmath,bm}
\usepackage{caption}
\usepackage{subcaption}
\usepackage{booktabs}   
\usepackage{multirow}
\usepackage{marvosym}

\usepackage{hyperref,xcolor}

\usepackage{acro}

\DeclareAcronym{lrp}{
  short = LRP ,
  long  = layerwise relevance propagation ,
  class = abbrev
}
\DeclareAcronym{lddmm}{
  short = LDDMM ,
  long  = large deformations diffeomorphic metric mapping ,
  class = abbrev
}
\DeclareAcronym{ssd}{
  short = SSD ,
  long  = sum of squared differences ,
  class = abbrev
}
\DeclareAcronym{mse}{
  short = MSE ,
  long  = mean squared error ,
  class = abbrev
}
\DeclareAcronym{mri}{
  short = MRI ,
  long  = magnetic resonance imaging ,
  class = abbrev
}
\DeclareAcronym{mr}{
  short = MR ,
  long  = magnetic resonance ,
  class = abbrev
}
\DeclareAcronym{tps}{
  short = TPS ,
  long  = thin-plate spline  ,
  class = abbrev
}
\DeclareAcronym{stn}{
  short = STN ,
  long  = spatial transformer network ,
  class = abbrev
  }
\DeclareAcronym{ga}{
  short = GA ,
  long  = gestational age ,
  class = abbrev
  }
\DeclareAcronym{pma}{
  short = PMA ,
  long  = post-menstrual age ,
  class = abbrev
  }
\DeclareAcronym{sam}{
  short = SAM ,
  long  = statistical appearance models ,
  class = abbrev
  }
\DeclareAcronym{cvae}{
  short = CVAE ,
  long  = conditional variational autoencoder  ,
  class = abbrev
  }
\DeclareAcronym{kl}{
  short = KL ,
  long  = Kullback-Leibler  ,
  class = abbrev
  }
\DeclareAcronym{ct}{
  short = CT ,
  long  = computed tomography ,
  class = abbrev
}
\DeclareAcronym{trus}{
  short = TRUS ,
  long  = transrectal ultrasound ,
  class = abbrev
}
\DeclareAcronym{us}{
  short = US ,
  long  = ultrasound ,
  class = abbrev
}
\DeclareAcronym{svf}{
  short = SVF ,
  long  = stationary velocity field ,
  class = abbrev
}
\DeclareAcronym{cnn}{
  short = CNN ,
  long  = convolutional neural network ,
  class = abbrev
}
\DeclareAcronym{fcn}{
  short = FCN ,
  long  = fully convolutional neural network ,
  class = abbrev
}
\DeclareAcronym{lstm}{
  short = LSTM ,
  long  = long short-term memory ,
  class = abbrev
}
\DeclareAcronym{cae}{
  short = CAE ,
  long  = convolutional autoencoder ,
  class = abbrev
}
\DeclareAcronym{t1w}{
  short = $T_1$w ,
  long  = $T_1$-weighted ,
  class = abbrev
}
\DeclareAcronym{t2w}{
  short = $T_2$w ,
  long  = $T_2$-weighted ,
  class = abbrev
}
\DeclareAcronym{dti}{
  short = DTI ,
  long  = diffusion tensor imaging ,
  class = abbrev
}
\DeclareAcronym{svd}{
  short = SVD ,
  long  = singular value decomposition ,
  class = abbrev
}
\DeclareAcronym{dt}{
  short = DT ,
  long  = diffusion tensor ,
  class = abbrev
}
\DeclareAcronym{dw}{
  short = DW ,
  long  = diffusion weighted ,
  class = abbrev
}
\DeclareAcronym{fodf}{
  short = fODF ,
  long  = fibre orientation distribution function ,
  class = abbrev
}
\DeclareAcronym{lncc}{
  short = LNCC ,
  long  = local normalised cross correlation ,
  class = abbrev
}
\DeclareAcronym{lcc}{
  short = LCC ,
  long  = local cross correlation ,
  class = abbrev
}
\DeclareAcronym{ncc}{
  short = NCC ,
  long  = normalised cross correlation ,
  class = abbrev
}
\DeclareAcronym{mi}{
  short = MI ,
  long  = mutual information ,
  class = abbrev
}
\DeclareAcronym{nmi}{
  short = NMI ,
  long  = normalised mutual information ,
  class = abbrev
}

\begin{document}
%
\title{Diffusion tensor driven image registration: a deep learning approach}

\titlerunning{Diffusion tensor driven image registration}



\author{Irina Grigorescu\inst{1}\textsuperscript{\Letter} \and 
Alena Uus\inst{1} \and
Daan Christiaens\inst{1}\inst{2} \and
Lucilio Cordero-Grande\inst{1} \and 
Jana Hutter\inst{1} \and
A. David Edwards\inst{1} \and 
Joseph V. Hajnal\inst{1} \and 
Marc Modat\inst{1} \and
Maria Deprez\inst{1}}

\authorrunning{I. Grigorescu et al.}
\institute{School of Biomedical Engineering \& Imaging Sciences, King's College London, London, UK \and Departments of Electrical Engineering, ESAT/PSI, KU Leuven, Leuven, Belgium \\
\textsuperscript{\Letter}\email{irina.grigorescu@kcl.ac.uk} }

\maketitle              
%
\begin{abstract}
Tracking microsctructural changes in the developing brain relies on accurate inter-subject image registration.
However, most methods rely on either structural or diffusion data to learn the spatial correspondences between two or more images, without taking into account the complementary information provided by using both.
Here we propose a deep learning registration framework which combines the structural information provided by \ac{t2w} images with the rich microstructural information offered by \ac{dti} scans.
This allows our trained network to register pairs of images in a single pass.
We perform a leave-one-out cross-validation study where we compare the performance of our multi-modality registration model with a baseline model trained on structural data only, in terms of Dice scores and differences in fractional anisotropy (FA) maps.
Our results show that in terms of average Dice scores our model performs better in subcortical regions when compared to using structural data only.
Moreover, average sum-of-squared differences between warped and fixed FA maps show that our proposed model performs better at aligning the diffusion data.
\keywords{image registration \and diffusion tensor imaging}
\end{abstract}

\section{Introduction}

Medical image registration is a vital component of a large number of clinical applications.
For example, image registration is used to track longitudinal changes occurring in the brain.
However, most applications in this field rely on a single modality, without taking into account the rich information provided by other modalities.
Although \ac{t2w} \ac{mri} scans provide good contrast between different brain tissues, they do not have knowledge of the extent or location of white matter tracts.
Moreover, during early life, the brain undergoes dramatic changes, such as cortical folding and myelination, processes which affect not only the brain's shape, but also the \ac{mri} tissue contrast.

In order to establish correspondences between images acquired at different gestational ages, we propose a deep learning image registration framework which combines both \ac{t2w} and \ac{dti} scans.
More specifically, we build a neural network starting from the popular diffeomorphic VoxelMorph framework \cite{Balakrishnan2019}, on which we add layers capable of dealing with \ac{dt} images.
The key novelties in our proposed deep learning registration framework are:
\begin{itemize}
    \item[$\bullet$] The network is capable of dealing with higher-order data, such as \ac{dt} images, by accounting for the change in orientation of diffusion tensors induced by the predicted deformation field.
    
    \item[$\bullet$] During inference, our trained network can register pairs of \ac{t2w} images without the need to provide the extra microstructural information.
    This is helpful when higher-order data is missing in the test dataset.
\end{itemize}

Throughout this work we use 3-D \ac{mri} brain scans acquired as part of the developing Human Connectome Project\footnote{\href{http://www.developingconnectome.org/}{http://www.developingconnectome.org/}} (dHCP).
We showcase the capabilities of our proposed framework on images of infants born and scanned at different gestational ages and we compare the results against the baseline network trained on only \ac{t2w} images.
Our results show that by using both modalities to drive the learning process we achieve superior alignment in subcortical regions and a better alignment of the white matter tracts.

\section{Method}

Let $F, \, M$ represent the fixed (target) and the moving (source) \ac{mr} volumes, respectively, defined over the 3-D spatial domain $\Omega$, and let $\phi$ be the deformation field.
In this paper we focus on \ac{t2w} images ($F^{T2w}$ and $M^{T2w}$ which are single channel data) and \ac{dt} images ($F^{DTI}$ and $M^{DTI}$ which are 6 channels data) acquired from the same subjects.
Our aim is to align pairs of \ac{t2w} volumes using similarity metrics defined on both the \ac{t2w} and \ac{dti} data, while only using the structural data as input to the network.

In order to achieve this, we model a function $g_{\theta}(F^{T2w}, M^{T2w}) = v$ a velocity field (with learnable parameters $\theta$) using a \ac{cnn} architecture based on VoxelMorph \cite{Balakrishnan2019}.
In addition to the baseline architecture, we construct layers capable of dealing with the higher-order data represented by our \ac{dt} images.
Throughout this work we use \ac{t2w} and \ac{dti} scans that have been affinely aligned to a common $40$ weeks gestational age atlas space \cite{Schuh2018}, prior to being used by the network.

\begin{figure}[ht]
\centering
\includegraphics[width=\textwidth,angle=0]{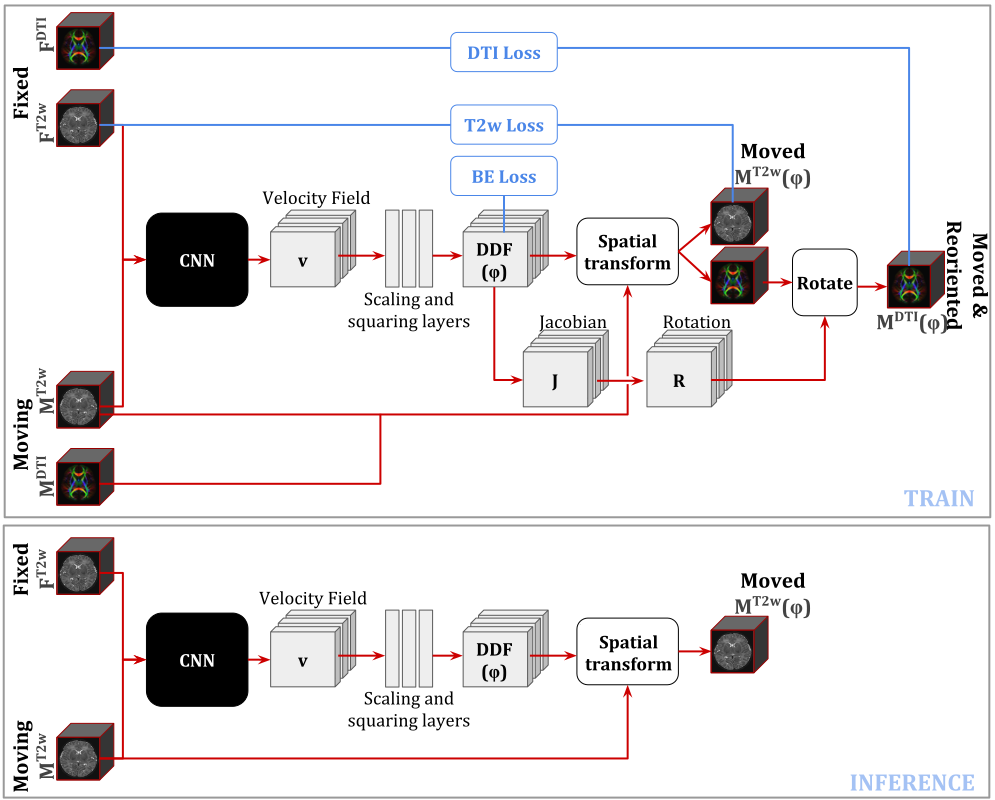}
\caption{The proposed network architecture at both training and inference time.}
\label{fig:networkArchitecture}
\end{figure}

Figure \ref{fig:networkArchitecture} shows the general architecture of the proposed network.
During training, our model uses pairs of \ac{t2w} images to learn a velocity field $v$, while the \textit{squaring and scaling layers} \cite{Balakrishnan2019} transform it into a topology-preserving deformation field $\phi$.
The moving images $M$ are warped by the deformation field using a \textit{SpatialTransform} layer \cite{Dalca2018} which outputs the moved (linearly resampled) \ac{t2w} and \ac{dt} images.
The \ac{dt} images are further processed to obtain the final moved and reoriented image.

The model is trained using stochastic gradient descent to find the optimal parameters $\hat{\theta}$ that minimize a sum of three loss functions, representing the tensor similarity measure, the scalar-data similarity measure and a regulariser applied on the predicted deformation field.
The \ac{dti} data is not used as input to our \ac{cnn}, but only used to drive the learning process through calculating the similarity measure.
During inference, our model uses only \ac{t2w} images to predict the deformation field, without the need for a second modality.
In the following subsections, we describe our model in further detail.







\paragraph{\textbf{Network architecture}}

The baseline architecture of our network is a 3-D UNet \cite{Ronneberger2015} based on VoxelMorph \cite{Balakrishnan2019}.
The encoding branch is made up of four 3D convolutions of $16, 32, 32$, and $32$ filters, respectively, with a kernel size of $3 \times 3 \times 3$, followed by \textit{Leaky ReLU} ($\alpha = 0.2$) activations \cite{xu2015empirical}.
The decoding branch contains four transverse 3D convolutions of 32 filters each, with the same kernel size and activation function.
Skip connections are used to concatenate the encoding branch's information to the decoder branch.
Two more convolutional layers, one with $16$ filters and a second one with $3$ filters, are added at the end, both with the same kernel size and activation function as before.

A pair of \ac{t2w} images are concatenated on the channel axis and become a $96 \times 96 \times 64 \times 2$ input for the \ac{cnn} network. 
The output is a three channel velocity field of the same size as the input images.
The velocity field is smoothed with a $3 \times 3 \times 3$ Gaussian kernel (with $\sigma = 1.2$ mm), and passed onto seven \textit{squaring and scaling layers} \cite{Balakrishnan2019}, which transform it into a topology-preserving deformation field.
The \textit{SpatialTransform} layer \cite{Dalca2018} receives as input the predicted field $\phi$ and the moving scalar-valued \ac{t2w} image, and outputs the warped and resampled image.
A similar process is necessary to warp the moving \ac{dt} image, with a few extra steps which are explained in the next subsection.

\paragraph{\textbf{Tensor reorientation}}

Registration of \ac{dt} images is not as straightforward to perform as scalar-valued data.
When transforming the latter, the intensities in the moving image are interpolated at the new locations determined by the deformation field $\phi$ and copied to the corresponding location in the target image space.
However, after interpolating \ac{dt} images, the diffusion tensors need to be reoriented to remain anatomically correct \cite{alexander2001spatial}.
In this work we use the \textit{finite strain} (FS) strategy \cite{alexander2001spatial}.

When the transformation is non-linear, such as in our case, the reorientation matrix can be computed at each point in the deformation field $\phi$ through a polar decomposition of the local Jacobian matrix.
This factorisation transforms the non-singular matrix $J$ into a unitary matrix $R$ (the pure rotation) and a positive-semidefinite Hermitian matrix $P$, such that $J = RP$ \cite{shoemake1992matrix}.
The rotation matrices $R$ are then used to reorient the tensors without changing the local microstructure.


\paragraph{\textbf{Loss function}}

We train our model using a loss function composed of three parts.
First, the structural loss $\mathcal{L}_{struct}$ (applied on the \ac{t2w} data only) is a popular similarity measure used in medical image registration, called \ac{ncc}.
We define it as:
\[ 
NCC(F,M(\phi)) = - \frac{\sum_{\mathbf{x} \in \Omega} (F(\mathbf{x}) - \overline{F}) \cdot (M(\phi(\mathbf{x})) - \overline{M})}{\sqrt{\sum_{\mathbf{x} \in \Omega} (F(\mathbf{x}) - \overline{F})^2 \cdot \sum_{\mathbf{x} \in \Omega} (M(\phi(\mathbf{x})) - \overline{M})^2}}
\]
where $\overline{F}$ is the mean voxel value in the fixed image $F$ and $\overline{M}$ is the mean voxel value in the transformed moving image $M(\phi)$.

Second, to encourage a good alignment between the \ac{dt} images, we set $\mathcal{L}_{tensor}$ to be one of the most commonly used diffusion tensor similarity measures, known as the Euclidean distance squared.
We define it as:
\[ EDS(F,M(\phi)) = \sum_{\mathbf{x} \in \Omega} \lvert \lvert F(\mathbf{x}) - M(\phi(\mathbf{x})) \rvert \rvert_C^2 
\] 
where the euclidean distance between two pairs of tensors $\mathbf{D_1}$ and $\mathbf{D_2}$ is defined as $\lvert \lvert \mathbf{D_1} - \mathbf{D_2} \rvert \rvert_C = \sqrt{ Tr ( ( \mathbf{D_1} - \mathbf{D_2})^2 ) }$  \cite{zhang2006deformable}.

Finally, to ensure a smooth deformation field $\phi$ we use a regularisation penalty $\mathcal{L}_{reg}$ in the form of bending energy \cite{Rueckert1999}:
\begin{flalign*}
    BE(\phi) = \sum_{\mathbf{x} \in \Omega} & \Big[ \Big(\frac{\partial^2 \phi(\mathbf{x})}{\partial x^2} \Big) ^2 + 
    \Big(\frac{\partial^2 \phi(\mathbf{x})}{\partial y^2} \Big) ^2 +
    \Big(\frac{\partial^2 \phi(\mathbf{x})}{\partial z^2} \Big) ^2 + \\
    &   
    2 \Big(\frac{\partial^2 \phi(\mathbf{x})}{\partial xy} \Big) ^2 +
    2 \Big(\frac{\partial^2 \phi(\mathbf{x})}{\partial xz} \Big) ^2 +
    2 \Big(\frac{\partial^2 \phi(\mathbf{x})}{\partial yz} \Big) ^2
    \Big]
\end{flalign*}

Thus, the final loss function is:
\begin{flalign*}
    \mathcal{L} (F, M(\phi)) = \alpha \, EDS(F^{DTI},M^{DTI}(\phi)) + \beta \, NCC(F^{T2w},M^{T2w}(\phi)) + \lambda \, BE(\phi)
\end{flalign*}

We compare our network with a baseline trained on \ac{t2w} data only.
For the latter case the loss function becomes: $\mathcal{L} (F, M(\phi)) = \beta \, NCC(F^{T2w},M^{T2w}(\phi)) + \lambda \, BE(\phi)$.
In all of our experiments we set the weights to 
$\alpha = 1.0$, $\beta = 1.0$ and $\lambda = 0.001$ when using both \ac{dti} and \ac{t2w} images,
and to $\beta = 1.0$ and $\lambda = 0.001$ when using \ac{t2w} data only.
These hyper-parameters were found to be optimal on our validation set.

\section{Experiments}

\paragraph{\textbf{Dataset}}

The image dataset used in this work is part of the developing Human Connectome Project.
Both the \ac{t2w} images and the \ac{dw} images were acquired using a 3T Philips Achieva scanner and a 32-channels neonatal head coil \cite{Hughes2017}.
The structural data was acquired using a turbo spin echo (TSE) sequence in two stacks of 2D slices (sagittal and axial planes), with parameters: $T_R = 12$ s, $T_E = 156$ ms, and SENSE factors of 2.11 for the axial plane and 2.58 for the sagittal plane. 
The data was subsequently corrected for motion \cite{CorderoGrande2018,kuklisova2012reconstruction} and resampled to an isotropic voxel size of $0.5$ mm. 

The \ac{dw} images were acquired using a monopolar spin echo echo-planar imaging (SE-EPI) Stejksal-Tanner sequence \cite{hutter2018time}.
A multiband factor of 4 and a total of 64 interleaved overlapping slices ($1.5$ mm in-plane resolution, $3$ mm thickness, $1.5$ mm overlap) were used to acquire a single volume, with parameters $T_R = 3800$ ms, $T_E = 90$ ms.
This data underwent outlier removal, motion correction and it was subsequently super-resolved to a $1.5$ mm isotropic voxel resolution \cite{christiaens2019scattered}.
All resulting images were checked for abnormalities by a paediatric neuroradiologist.

\begin{figure}
\centering
\includegraphics[width=.7\textwidth]{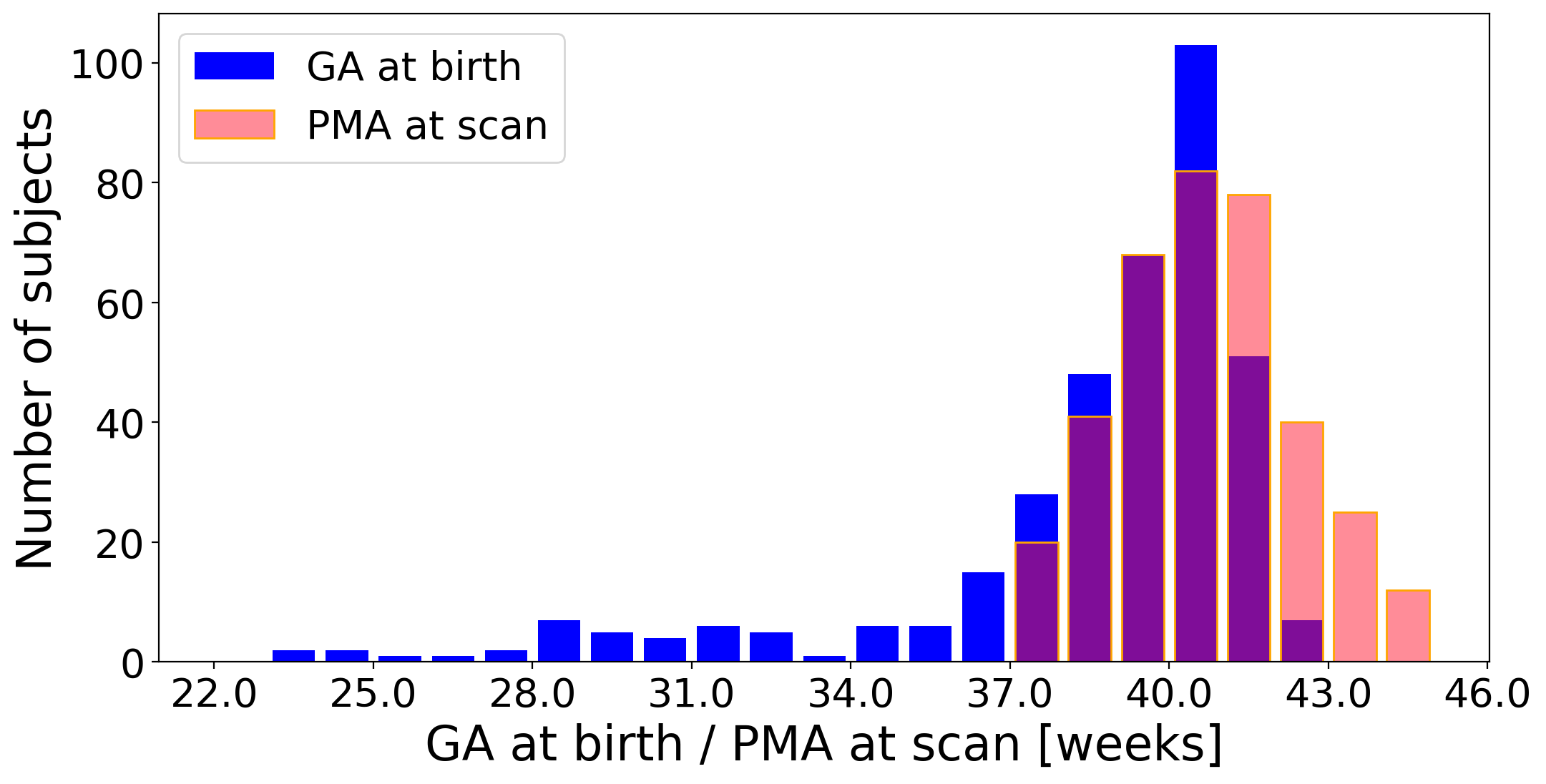}
\caption{Distribution of gestational ages at birth (GA) and post-menstrual ages at scan (PMA) in our dataset.} 
\label{fig:distribBabies}
\end{figure}

For this study, we use a total of 368 \ac{t2w} and \ac{dt} volumes of neonates born between $23 - 42$ weeks \ac{ga} and scanned at term-equivalent age ($37-45$ weeks \ac{ga}).
The age distribution in our dataset is found in Figure~\ref{fig:distribBabies}, where \ac{ga} at birth is shown in blue, and \ac{pma} at scan is shown in orange.
In order to use both the \ac{t2w} and \ac{dt} volumes in our registration network, we first resampled the \ac{t2w} data into the \ac{dw} space of $1.5$ mm voxel resolution.
Then, we affinely registered all of our data to a common 40 weeks gestational age atlas space \cite{Schuh2018} available in the MIRTK\footnote{\href{https://mirtk.github.io/}{https://mirtk.github.io/}} software toolbox \cite{Rueckert1999} and obtained the \ac{dt} images using the \textsc{dwi2tensor} \cite{veraart2013weighted} command available in the MRTRIX\footnote{\href{https://mrtrix.readthedocs.io/}{https://mrtrix.readthedocs.io/}} toolbox.
Finally, we performed skull-stripping using the available dHCP brain masks \cite{christiaens2019scattered} and we cropped the resulting images to a $96 \times 96 \times 64$ volume size. 

\paragraph{\textbf{Training}}

We trained our models using the rectified Adam (RAdam) optimiser \cite{liu2019variance} with a cyclical learning rate \cite{smith2015cyclical} varying from $10^{-9}$ to $10^{-4}$, for $90,000$ iterations.
Out of the 368 subjects in our entire dataset, 318 were used for training, 25 for validation and 25 for test.
The subjects in each category were chosen such that their \ac{ga} at birth and \ac{pma} at scan were distributed across the entire range.
The validation set was used to help us choose the best hyperparameters for our network and the best performing models.
The results reported in the next section are on the test set.

\paragraph{\textbf{Final model results}}

\begin{figure}
     \centering
     \includegraphics[width=\textwidth,angle=0]{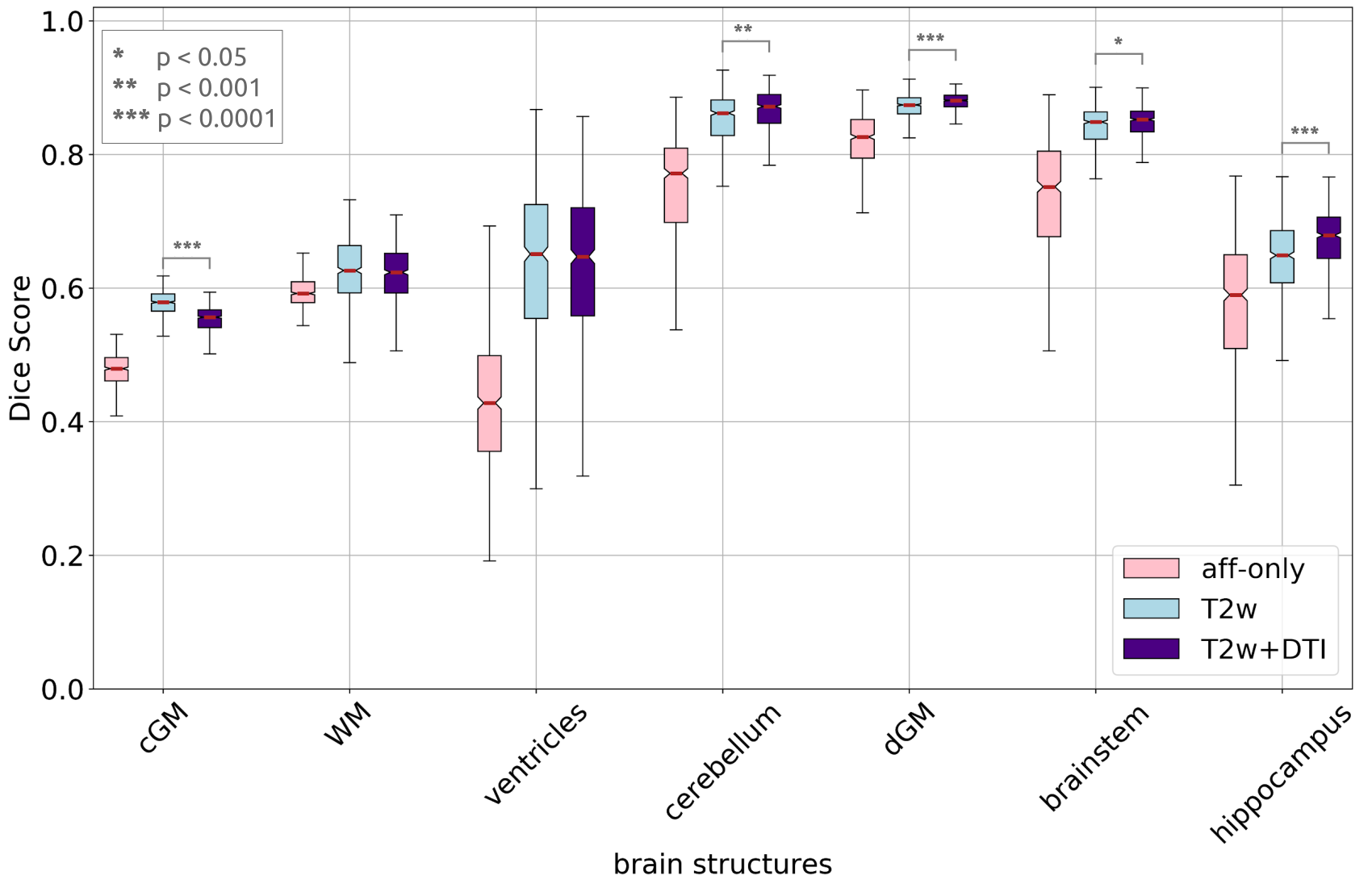}
    \caption{Average Dice scores for our cross-validation study for 7 tissue types: cortical gray matter (cGM), white matter (WM), ventricles, cerebellum, deep gray matter (dGM), brainstem and the hippocampus. 
    For both of our trained models the input images, $F^{T2w \lvert DTI}$ and $M^{T2w \lvert DTI}$, have been affinely aligned to a template, prior to being used by the models.
    Our proposed model outperforms the \ac{t2w}-only training in terms of obtaining higher Dice scores for the cerebellum, dGM, brainstem and hippocampus.}
    \label{fig:dicescores}
\end{figure}

In both our \ac{t2w}-only and \ac{t2w}+\ac{dti} cases we performed a leave-one-out cross-validation, where we aligned 24 of the test subjects to a single subject, and repeated until all the subjects were used as target.
Each of the 25 subjects had tissue label segmentations (obtained using the Draw-EM pipeline for automatic brain MRI segmentation of the developing neonatal brain \cite{makropoulos2014automatic}) which were propagated using \textsc{NiftyReg}\footnote{\href{https://github.com/KCL-BMEIS/niftyreg/}{https://github.com/KCL-BMEIS/niftyreg/}} \cite{modat2010fast} and the predicted deformation fields.
The average resulting Dice scores are summarised in Figure \ref{fig:dicescores}, where the initial pre-alignment is shown in pink, the \ac{t2w}-only results are shown in light blue and the \ac{t2w}+\ac{dti} are shown in purple.
Our proposed model performs better than the baseline model for all subcortical structures (cerebellum, deep gray matter, brainstem and hippocampi and amygdala), while performing similarly well in white matter structures.
In contrast, cortical gray matter regions were better aligned when using the \ac{t2w}-only model, as structural data has higher contrast than DTI in these areas.


We also computed the FA maps for all the initial affinely aligned and all the warped subjects in the cross-validation study and calculated the sum-of-squared differences (SSD) between the moved FA maps and the fixed FA maps.
The resulting average values are summarised in Table~\ref{tab:ssdfa}, which shows that our proposed model achieved better alignment in terms of FA maps.

\begin{table}
\begin{center}
 \begin{tabular}{c c c | c c c r } 
 \textbf{Method} & \, \textbf{Mean(SSD)} & \, \textbf{Std.Dev.(SSD)} & \multicolumn{4}{c}{\textit{\textbf{p-value}}} \\ [0.5ex] 
 \hline
  Affine & 1087 & 174 & \, Affine & vs & T2w & \, $p < 1e^{-5}$ \\ 
  T2w & 1044 & 168 & \, Affine & vs & T2w+DTI & \,  $p < 1e^{-5}$ \\ 
  \textbf{T2w+DTI} & 981 & 181 & \, T2w & vs & T2w+DTI & \, $p < 1e^{-5}$ \\
\end{tabular}
\end{center}
\caption{Average sum-of-squared differences between warped and fixed FA maps in our leave-one-out cross-validation study. The first line shows mean and standard deviation SSD values for the initial affine alignment.}
\label{tab:ssdfa}
\end{table}

\begin{figure}
     \centering
    \includegraphics[width=1\textwidth,angle=0]{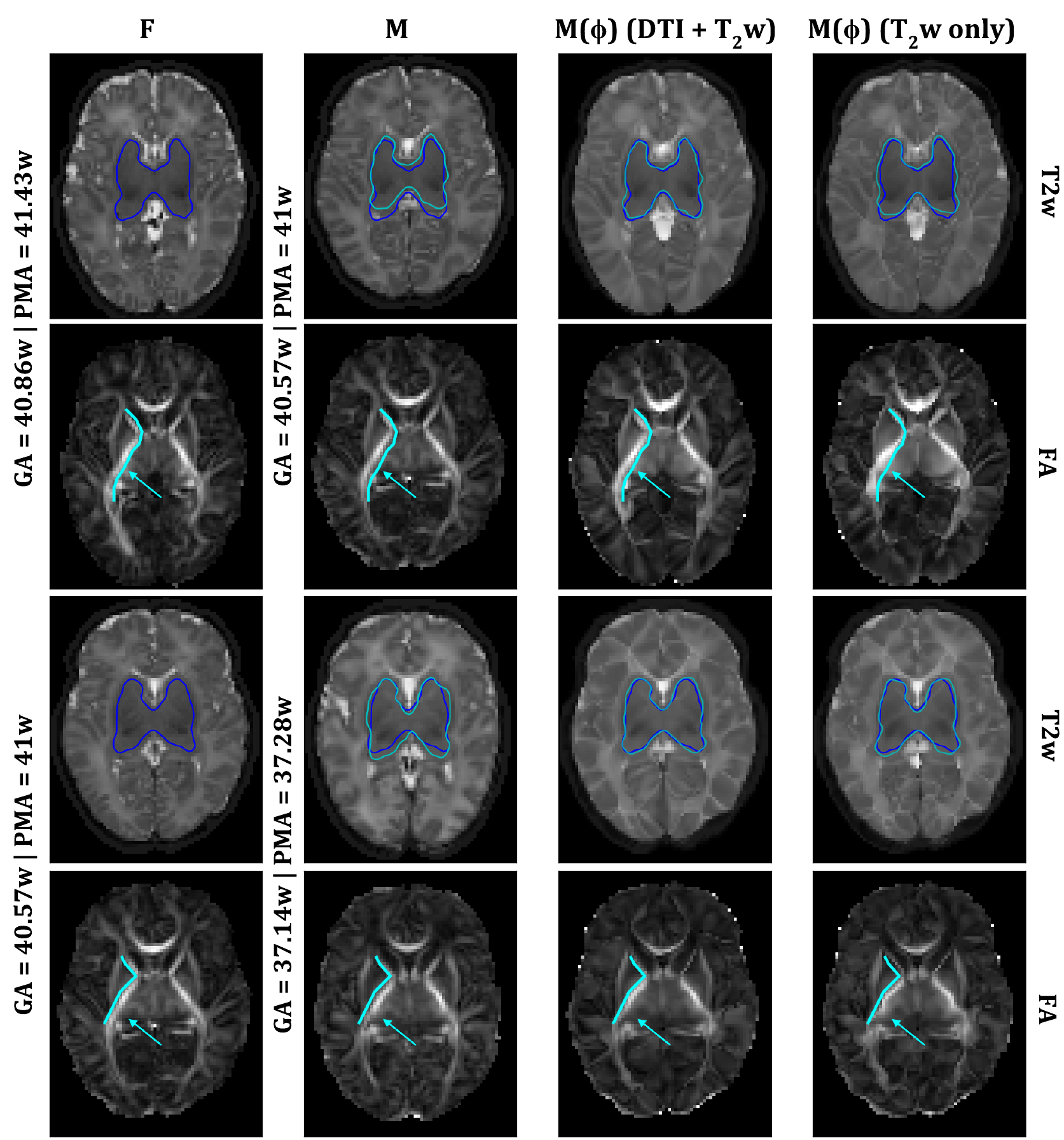}
    \caption{First two rows show an example registration between a neonate with GA = 40.57w and PMA = 41w as moving, and one with GA = 40.86w and PMA = 41.43w as fixed, last two rows show an example where the moving image is from a neonate with GA = 37.14w and PMA = 37.28w, and fixed is a neonate with GA = 40.57w and PMA = 41w.
    First column shows axial slices of the fixed \ac{t2w} images and FA maps, the second column shows axial slices of the moving \ac{t2w} images and FA maps, and the third and fourth columns show the moved images using our proposed network and the baseline network, respectively.
    In the \ac{t2w} maps the deep gray matter (dGM) labels are shown for the fixed images in dark blue and for the moving and moved in cyan.
    In both cases a higher dGM Dice score was obtained for the \ac{t2w}+\ac{dti} model (0.88 and 0.88, respectively), than when using \ac{t2w}-only (0.84 and 0.87, respectively).
    The arrows point at areas where the underlying anatomy was better preserved when using \ac{t2w}+\ac{dti}, than when using \ac{t2w}-only.}
    \label{fig:exampleRegistration}
\end{figure}

Finally, Figure~\ref{fig:exampleRegistration} shows two example registrations.
The target images are from two term-born infants with GA = 40.86 weeks and PMA = 41.43 weeks, and GA = 40.57w and PMA = 41w, respectively, while the moving images are from infants with GA = 40.57 weeks and PMA = 41 weeks, and GA = 37.14w and PMA = 37.28w, respectively.
The figure shows both \ac{t2w} and FA maps of axial slices of the fixed (first column), the moving (second column) and the warped images by our proposed method (third column) and the baseline method (fourth column), respectively.
The moved FA maps show that by using \ac{dti} data to drive the learning process of a deep learning registration framework, we were able to achieve good alignment not only on the structural data, but also on the diffusion data as well.

\section{Discussion and future work}

In this work we showed for the first time a deep learning registration framework capable of aligning both structural (T2w) and microstructural (DTI) data, while using only \ac{t2w} data at inference time.
A key result from our study is that our proposed \ac{t2w}+\ac{dti} model performed better in terms of aligning subcortical structures, even though the labels for these regions were obtained from structural data only.
For future work we plan to focus on improving the registration in the cortical regions, and to compare our deep learning model with classic registration algorithms.



\section*{Acknowledgments}

This work  was supported by the Academy of Medical Sciences Springboard Award (SBF004\textbackslash1040), European Research Council under the European Union’s Seventh Framework Programme (FP7/ 20072013)/ERC grant agreement no. 319456 dHCP project, the Wellcome/EPSRC Centre for Medical Engineering at King’s College London (WT 203148/Z/16/Z), the NIHR Clinical Research Facility (CRF) at Guy's and St Thomas' and by the National Institute for Health Research Biomedical Research Centre based at Guy's and St Thomas' NHS Foundation Trust and King’s College London. The views expressed are those of the authors and not necessarily those of the NHS, the NIHR or the Department of Health.

%
%
\bibliographystyle{splncs04}
\bibliography{bibliographywbir}

\end{document}